\documentclass[11pt,preprint]{aastex}
\newcommand{\Msol}{M$_{\odot}$}
\newcommand{\Msun}{M$_{\odot}$}

\newcommand{\Mjup}{M$_{\mathrm{JUP}}$}

\newcommand{\Msini}{M\,$\sin i$}

\newcommand{\ms}{m\,s$^{-1}$}

\newcommand{\rhk}{R$^{\prime}_{HK}$}

\newcommand{\be}{\begin{equation}}
\newcommand{\ee}{\end{equation}}

\received{December 2005}
\accepted{February 2006s}
\slugcomment{To appear in  {\em The Astrophysical Journal}.}

\shortauthors{Tinney et al.}
\shorttitle{2:1 Resonant Exoplanetary System HD73526}

\begin{document}
\title{The 2:1 resonant exoplanetary system orbiting HD73526\altaffilmark{1} }

\author{C.G. Tinney\altaffilmark{2}, R. Paul Butler\altaffilmark{3}, 
        Geoffrey W. Marcy\altaffilmark{4,5}, Hugh R.A. Jones\altaffilmark{6}, 
        Gregory Laughlin\altaffilmark{7},
        Brad D. Carter\altaffilmark{8}, Jeremy A. Bailey\altaffilmark{9},
        Simon O'Toole\altaffilmark{2}}

\altaffiltext{1}{Based on observations obtained at the
                 Anglo--Australian Telescope, Siding Spring, Australia.}
\altaffiltext{2}{Anglo-Australian Observatory, PO Box 296, Epping. 1710. 
                 Australia. {\tt cgt@aaoepp.aao.gov.au}}
\altaffiltext{3}{Carnegie Institution of Washington,Department of Terrestrial Magnetism,
                 5241 Broad Branch Rd NW, Washington, DC 20015-1305}
\altaffiltext{4}{Department of Astronomy, University of California, Berkeley, CA, 94720}
\altaffiltext{5}{Department of Physics and Astronomy, 
                 San Francisco State University, San Francisco, CA 94132.}
\altaffiltext{6}{Centre for Astrophysics Research, University of Hertfordshire, 
                 Hatfield AL10 9AB UK}
\altaffiltext{8}{UCO/Lick Observatory, University of California at Santa Cruz, 
                 Santa Cruz, CA 95064.}
\altaffiltext{8}{Faculty of Sciences, University of Southern Queensland, 
                 Toowoomba, 4350. Australia.}
\altaffiltext{9}{Australian Centre for Astrobiology, Macquarie University. 
                 2109. Australia.}

\begin{abstract}
We report the detection of a second exoplanet orbiting the G6V dwarf HD73526.
This second planet has an orbital period of 377\,d, putting it in a 2:1 
resonance with the previously known exoplanet, the orbital period for which
is updated to 188\,d. Dynamical modeling 
of the combined system allows solution for a self-consistent set of
orbital elements for both components.
HD\,73526 is the fourth exoplanetary system (of a total of
18 systems with 2 or more components currently known) to have components
detected in 2:1 resonance. Finding such a large fraction of multiple planets 
(more than 20 per cent) in 2:1 resonance strongly suggests
that orbital migration, halted by stabilisation in a trapping resonance, 
plays an important role in the evolution of exoplanets in multiple planet
systems.

\end{abstract}


\section{Introduction}

The Anglo-Australian Planet Search (AAPS) began taking data in January 1998 on the
nearest and brightest Sun--like stars. Results from this programme
\citep{AAPSI,AAPSIII,AAPSVII,AAPSXI,AAPSII,AAPSV,AAPSIV,AAPSVI,AAPSVIII, 
AAPSIX,AAPSX}  have demonstrated long-term velocity precisions of 3\,\ms\ or
better, for suitably quiescent Sun-like stars. The AAPS, together with programmes
using similar techniques on the Lick 3\,m and Keck 10\,m telescopes
\citep{fischer01,vmba00},  provides all-sky planet search coverage for inactive
F,G,K and M dwarfs out to distances of 50\,pc. For recent reviews of the progress
in exoplanetary detection and the Doppler technique, the interested reader is
referred to \citet{marcy05b} and \citet{mayor2005}, and references therein.

AAPS is being carried out on the 3.9\,m Anglo-Australian Telescope (AAT), using the
University College London Echelle Spectrograph (UCLES) and an I$_2$ absorption
cell.  UCLES is operated in its 31\,lines\,mm$^{-1}$ mode. Prior to 2001 September
it was used with  a MIT/LL 2048$\times$4096 15$\mu$m pixel CCD, and since then has
been used with an EEV 2048$\times$4096 13.5$\mu$m pixel CCD.  AAPS currently
observes on 32 nights per year.  The survey initially targeted 200 F,G,K and M
stars with $\delta < -20$\arcdeg\ and V$<$7.5. Where age/activity information was
available from \rhk\ indices (see for example \citealt{hsdb96,CaHKI}) we required
target stars to have $\log$\rhk $>$ $-$4.5, corresponding to ages greater than
3\,Gyr.

In addition to the primary sample,  a small sub-sample of twenty fainter dwarfs
with $uvby$ photometry  suggesting metal-enrichment over solar, was added in
October 1999 \citep{AAPSVII}.  These dwarfs have V$<$9 and were added to examine
suggestions that metal-enriched stars preferentially host planets (see e.g..
\citealt{laughlin00} and references therein).  HD\,73526 (the primary focus of this
paper) was one of these stars. A further 8 M-dwarfs extending as faint as V$<$11
were also included in  the program. In 2002, AAPS further expanded the scope of its
survey, increasing from 20 nights/year to 32 nights/year.  Sixty new stars were
then added to the target list, and those stars found since their initial inclusion
to have \rhk\ activity levels inconsistent with high precision velocity measurement
(i.e.. $\log$\rhk $<$ $-$4.5 \citealt{CaHKI,CaHKII}) were culled.  
The resulting AAPS target sample
currently includes 253 stars. Our observing procedure and data analysis  continue
to substantially follow that described in \citet{bmwmd96} and  \citet{AAPSII},
though over the last 24 months significant improvements have been made to the
spectral extraction component of the analysis package \citep{AAPSXI}. 
A signal-to-noise ratio of at least 200 per pixel is now
standard for all stars.

The detection of a first extra-solar planet orbiting the star HD\,73526 was
presented in \citet{AAPSVII}. This planet (hereafter HD\,73526b) was estimated to
have an orbital period $P = $190.5$\pm$3.0\,d, eccentricity $e = $0.34$\pm$0.08 and
amplitude $K = $108$\pm$8\,\ms, leading to a minimum mass estimate of \Msini $=$
3.0$\pm$0.3\Mjup.

%
%

\section{The star HD73526}

%

%

%
%

%

%
%

The characteristics of the host star HD73526 are summarised in Table
\ref{parameters} -- please refer to the table notes for references. HD\,73526
(HIP\,42282, SAO\,220191) is a G6V dwarf. No \rhk\ estimate is currently available.
HIPPARCOS finds it to be photometrically stable. HD\,73526 was initially added to
our AAPS target sample based on Str\"omgren $uvby$ photometry  suggestion metal
enrichment over solar, and based on which  \citet{AAPSVII} estimated a metallicity
in the range $[$Fe/H$]$=0.10 to 0.16 (see also \citealt{nord2004} and references
therein). Several independent detailed spectroscopic analyses have now been
performed for  this star leading to metallicity estimates of
$[$Fe/H$]$=$+$0.25$\pm$0.03 \citep{fv05}, $+$0.27$\pm$0.06 \citep{santos2004},  and
$+$0.11$\pm$0.07 \citep{Bond05}.  From these we conclude that the metallicity of this
star is  somewhat higher than that derived from Str\"omgren photometry, and
assume a metallicity of $[$Fe/H$]$=$+$0.25. The effective temperatures estimated
by these various studies range from 5470K to 5700K, and we have assumed a median
value of 5590K. Both \citet{santos2004} and
\citet{fv05} have used isochrone interpolation to estimate a mass for this star,
and derive 1.05\Msun and 1.08\Msun (respectively).
We have adopted a mass of 1.08\Msun, giving more weight to the \citet{fv05} estimate
which is based on more recent isochrones.
These parameters are not significantly different from those
assumed by \citet{AAPSVII}, and continue to suggest that HD\,73526 is beginning its
evolution off the main sequence.

\section{Kinematic Orbital Solutions}
\label{velocities}

Thirty observations of HD\,73526 are listed in Table \ref{vel73526}. The  column
labeled ``Unc'' is the velocity uncertainty produced by the least-squares
fitting procedure. This fit simultaneously determines the Doppler shift and the
spectrograph point-spread function (PSF) for each observation made though the
iodine cell, given an iodine absorption spectrum and an ``iodine free'' template
spectrum of the object \citep{bmwmd96}.  The uncertainty is derived for each
measurement by taking the mean of four hundred useful spectral regions (each
2\,\AA\ long) from each exposure. This uncertainty includes the effects of
photon-counting uncertainties, residual errors in the spectrograph PSF model,
and variation in the underlying  spectrum between the template and ``iodine''
epochs. Since the internal velocity uncertainties produced by the least-squares
fitting procedure do not reflect the likely intrinsic
variability, or ``jitter'' of HD73526, a
nominal 3.3\,\ms jitter uncertainty is added in quadrature to these internal
velocity uncertaintes to use in generating the reduced chi-squared ($\chi_{\nu}^2$).
All velocities are measured relative to the  zero-point defined by the
template observation.  
\clearpage
%
\begin{deluxetable}{lccccc}
\tablenum{1}
\tablecaption{Stellar Parameters for HD\,73526}
\label{parameters}
\tablewidth{0pt}
\tablehead{
\colhead{Parameter}&\colhead{Value}&\colhead{References}}   
\startdata
HIPPARCOS  N$_{\mathrm obs}$\qquad\qquad  & 137  & 1 \\             
HIPPARCOS  $\sigma$          & 0.02              & 1 \\             
HIPPARCOS  $\pi$ (mas)       & 10.6$\pm$1.0      & 1 \\             
M$_{\mathrm V}$              & 4.1$\pm$0.2       & 1 \\             
M$_{\mathrm Bol}$            & 3.7$\pm$0.2       & 2 \\             
Spectral Type                & G6V               & 3 \\
$[$Fe$/$H$]$ (spec)          &$+$0.25$\pm$0.05   & 4 \\ 
T$_{\mathrm eff}$ (K)        &5590               & 4 \\ 
Mass (\Msol)                 &1.08$\pm$0.05      & 5 \\ 
\enddata
\tablenotetext{1}{\citet{esa97}}
\tablenotetext{2}{\citet{allenIV}}
\tablenotetext{3}{\citet{mssII}}
\tablenotetext{4}{\citet{santos2004, fv05, Bond05} -- see text}
\tablenotetext{5}{\citet{santos2004, fv05} -- see text}
\end{deluxetable}

%
\begin{deluxetable}{rrr}
\tablenum{2}
\tablecaption{Velocities for HD\,73526}
\label{vel73526}
\tablewidth{0pt}
\tablehead{
JD & RV & Unc \\
(-2450000)   &  (\ms) & (m s$^{-1}$)
}
\startdata
       1212.1302   &  -2.1  &   10.5 \\
       1213.1315   &   7.7  &   10.5 \\
       1214.2390   &   2.8  &   12.5 \\
       1236.1465   &   4.0  &   12.8 \\
       1630.0280   &   0.0  &    9.8 \\
       1717.9000   &-180.8  &   13.0 \\
       1920.1419   & -82.4  &   12.4 \\
       1984.0378   &   8.2  &    9.0 \\
       2009.0976   &  10.0  &    8.3 \\
       2060.8844   &-106.7  &    7.8 \\
       2091.8465   &-221.2  &   13.8 \\
       2386.9003   &  -3.8  &    6.3 \\
       2387.8921   &  -1.7  &    5.2 \\
       2420.9248   & -65.4  &    6.6 \\
       2421.9199   & -68.9  &    6.0 \\
       2422.8602   & -71.5  &    6.1 \\
       2424.9237   & -77.4  &   12.4 \\
       2454.8526   &-154.8  &    6.6 \\
       2655.1519   & -79.4  &    6.7 \\
       3008.1339   &   3.4  &    4.5 \\
       3045.1355   & -96.9  &    6.1 \\
       3399.1625   & -54.7  &    5.6 \\
       3482.8801   &  20.6  &    3.8 \\
       3483.8871   &  28.7  &    4.8 \\
       3485.9622   &  21.0  &    6.6 \\
       3488.9389   &   7.2  &    4.2 \\
       3506.8863   &   3.0  &    4.2 \\
       3508.9119   &  14.5  &    3.8 \\
       3515.8937   &  -1.8  &    5.6 \\
       3520.9103   &  -3.9  &    6.3 \\
\enddata
\tablenotetext{a}{Julian Dates (JD) are barycentric. Radial Velocities
(RV) are barycentric, but have an arbitrary zero-point determined by
the radial velocity of the template.}
\end{deluxetable}
\clearpage

In \citet{AAPSVII} it was noted that the rms residuals of 18\ms\ about the best
Keplerian fit to the HD73526 velocity data were significantly higher than
usually seen from AAPS data, even when the relative faintness of this star is
taken into account. Indeed the velocity uncertainties produced by our
least-squares fitting process were typically at half this level, and the 
$\chi_{\nu}^2$=1.63 for this fit was significantly above the
expected value of 1.0.

\begin{figure}
\centering\includegraphics[angle=90,width=90mm]{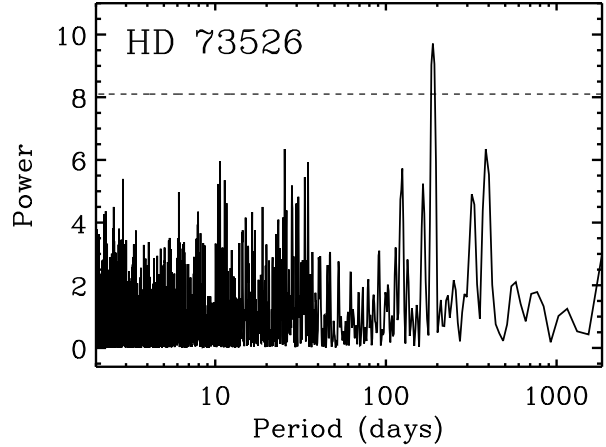} 
\caption{Power spectrum of velocities shown in Table \ref{vel73526}. The
horizontal dotted line represents a false alarm probability level of 1\%. The
three most significant peaks in this power spectrum are near 190\,d, 380\,d and
128\,d.}
\label{power}
\end{figure}

\begin{figure}
\centering\includegraphics[angle=90,width=90mm]{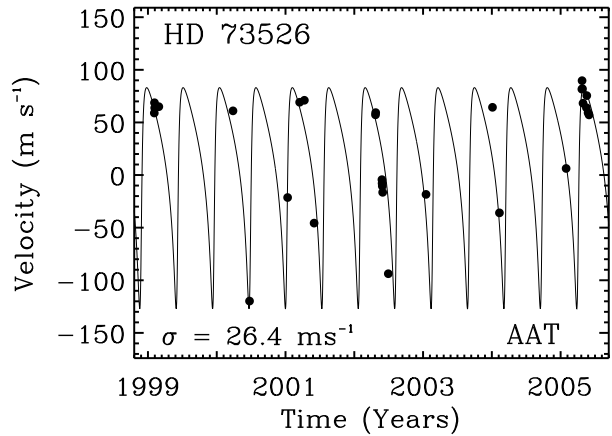}
\caption{Measured velocities of HD\,73526.  A single Keplerian fit to the data
with period 187.5d is plotted
over the data.  The rms residuals to this fit (26.4\ms) are very large, which
suggests this fit does not adequately parametrise the system. (Only  epochs from Table \ref{vel73526}  with internal uncertainty
less than twice the value of the median internal uncertainty are included.) } 
\label{hd73526_b_curve}
\end{figure}

Figure \ref{power} shows a power spectrum generated from the velocities in Table
\ref{vel73526}, indicating the presence of a strong peaks in observed power 
near the 190\,d period first detected by \citet{AAPSVII},  and additional peaks
(at increasing false  alarm probabilities) near 380\,d and 128\,d. Figure
\ref{hd73526_b_curve} shows the results for a best fit single Keplerian  period
near  190\,d.  As
suggested by the large residuals in our initial fit to the earlier set of data,
this single planet fit is clearly inadequate to correctly model  this data,
leading us to conclude that a multiple Keplerian fit is  required. This
situation led \citet{gregory2005}, in an independent reanalysis of the eighteen
epochs published by \citet{AAPSVII}, to suggest the likelihood  of other
periodicities near 128\,d and 376\,d in that data.

Adopting the set of parameters fitted for a single Keplerian
near the dominant 190\,d period as a likely neighborhood
for the orbital parameters
of a first planet, the data was searched for a subsequent planet by
performing dual Keplerian fits to the data with these first planet
parameters, and a period for the second planet selected as:
  (a)  the four highest peaks in the periodogram of the residuals
        to the first planet model model;
  (b)  twenty orbital periods spaced in equal logarithmic intervals,
       with 10 being less than and 10 being greater than
       the period of the first planet; and
  (c)  three periods that are twice, three times, and four times the
       highest periods found in (a) and (b), as trials for a planet
       with period much longer than the duration of
        observations.
For each of these 27 guesses for the period of the second planet,
a search of the vicinity is done to find a minimum of $\chi_{\nu}^2$.
The use of periodagram peaks and the logarithmic spacing of trial periods
will catch second planets as a minimum in $\chi_{\nu}^2$.
This process revealed that a 190\,d + 380\,d system clearly demonstrated
a minimal $\chi_{\nu}^2$. To reinforce this conclusion, we show in 
Table \ref{pairwise} the $\chi_{\nu}^2$ for dual Keplerian fits to all the
data epochs, using starting periods at pair-wise choices from the
three dominant periodogram peaks -- 190\,d, 380\,d and 128\,d. 

\clearpage
\begin{deluxetable}{lcccc}
\tablenum{3}
\tablecaption{HD\,73526b,c Possible Dual Keplerians}
\label{pairwise}
\tablewidth{0pt}
\tablehead{
\colhead{Starting Periods}\
                 & \colhead{Fit Period 1}
                                 & \colhead{Fit Period 1}
                                                 &\colhead{Reduced $\chi^2$} 
                                                       &\colhead{rms}
                                                                           }
\startdata
128\,d + 190\,d  & 125.7$\pm$0.1 & 186.1$\pm$0.2 & 2.7 &  9.7\ms \\
128\,d + 380\,d  & 125.3$\pm$0.2 & 384.5$\pm$2.1 & 2.4 &  8.9\ms \\
190\,d + 380\,d  & 187.3$\pm$0.5 & 376.8$\pm$1.2 & 1.07&  7.3\ms \\
\enddata
\end{deluxetable}
\clearpage

\begin{figure}
\centering\includegraphics[angle=90,width=90mm]{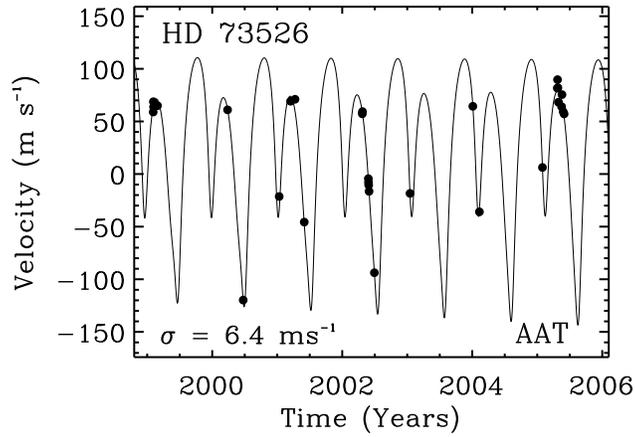}
\caption{Measured velocities of HD\,73526.  A double Keplerian fit to the data
with periods 187.5d and 376.9d and the parameters shown in Table
\ref{kin_orbits} is plotted over the data.  The rms residuals to this fit are
6.4\ms, and the reduced $\chi^2$=1.09, indicating the residuals about the fit
are consistent with measurement uncertainties. (Only  epochs from Table
\ref{vel73526}  with internal uncertainty less than twice the value of the
median internal uncertainty are included.)} 
\label{hd73526_bc_curve} 
\end{figure}

Figure \ref{hd73526_bc_curve} shows our best fitting dual Keplerian model. The
orbital parameters of this  solution are listed in Table \ref{kin_orbits}. Only 
epochs from Table \ref{vel73526}  with internal uncertainty less than twice the
value of the median internal uncertainty are included. The uncertainties in the
kinematic orbital parameters are derived from simulations as follows
\citep{marcy05b}. The set of residuals about the best-fit Keplerians are treated
as a population of random deviations with a  distribution characteristic of the
noise in the data.  We then randomly redistributed this ``noise'' onto
velocities calculated from  the best-fit solution at the observation epochs, and
refit to re-determine the orbital parameters. The uncertainties reported in
Table \ref{kin_orbits} are the standard deviations for each parameter that
result from repeating this procedure fifty times. 

This fit  suggests that the system is in a 
2:1 resonance with periods of 187.5\,d and 377\,d. 
It should be noted, however, that due to the similarity of the period of 
HD\,73526c to the orbital period of the Earth (and the consequent annual lack of
phase coverage when HD\,73526 passes behind the Sun), there is some degeneracy
between eccentricity and amplitude in our detailed solution  for the orbits of this
system. This degeneracy is not reflected in the uncertainties quoted in
the table, which uses our observation epochs as a basic assumption. While we can be confident of the orbital periods  of HD\,73526's two
planets,  equally valid solutions (as measured by $\chi^2$) can be derived with
eccentricities and amplitudes ranging from $e_b$,$e_c$=0.12,0.33 and
$K_b$,$K_c$=141,96 to $e_b$,$e_c$=0.36,0.38 and $K_b$,$K_c$=76,67 (where the
families of solutions trade off lower eccentricities for large amplitudes). This
state of affairs will change as monitoring of this system continues, though it will
take some  years for full phase coverage to be accessible with HD\,73526c's orbital
phase advancing at just 12\,d$/$376\,d = 3.2\% per year.

%

\clearpage
%
\begin{deluxetable}{lcccc}
\tablenum{4}
\tablecaption{HD\,73526b,c Kinematic Orbital Parameters}
\label{kin_orbits}
\tablewidth{0pt}
\tablehead{
\colhead{Parameter}            & \colhead{HD\,73526b}
                                                   & \colhead{HD\,73526c}
                                                                           }
\startdata
Orbital period $P$ (d)          &   187.5$\pm$0.3    & 376.9$\pm$0.9       \\
Velocity amp. $K$ (\ms)         &      76$\pm$5      &  67$\pm$4           \\
Eccentricity $e$                &    0.39$\pm$0.05   &  0.40$\pm$0.05      \\
$\omega$ (\arcdeg)              &    172$\pm$11      & 183$\pm$30          \\
$a_1 \sin i$ (km)               &  (180$\pm$4)$\times$10$^3$           
                                                     & (318$\pm$8)
                                                        $\times$10$^3$     \\
Periastron Time (JD-245000)     &   37$\pm$15        &  184$\pm$33         \\
\Msini\ (\Mjup)                 &  2.07$\pm$0.16     &  2.30$\pm$0.17       \\
a (AU)                          &  0.66$\pm$0.05     &  1.05$\pm$0.08       \\
$\chi_{\nu}^2$       & \multicolumn{2}{c}{1.09} \\
RMS (\ms)            & \multicolumn{2}{c}{6.4}\\
\enddata
\end{deluxetable}
\clearpage

\clearpage

\section{Dynamical Analysis}

The parameters listed in Table \ref{kin_orbits} were derived under the
approximation that the orbits are fixed Keplerian ellipses. In reality, the star
and the two  planets constitute an interacting three-body system. If one
interprets  the Keplerian parameters as osculating orbital elements
corresponding to a particular epoch, there is no guarantee that the resulting
configuration of masses is either dynamically stable, or consistent with the
radial velocity data set. Indeed, in this particular case, when the dual
Keplerian orbital elements listed in Table \ref{kin_orbits} are integrated
forward in time,  the system becomes unstable within a few thousand years.
Furthermore, when the system is integrated  from JD 2451212.1302, the epoch of
the first radial velocity data point,  the $\chi^{2}_{\nu}$ for the fit
increases from $\chi^{2}_{\nu}=1.4$ to $\chi^{2}_{\nu}=68.9$, and the RMS 
scatter increases
from 7.3\,\ms to 35.6\,\ms. (In these fits and
in the following dynamical analysis, all data epochs are used and
the uncertainties used are the straight
internal uncertainties from Table \ref{vel73526} without additional jitter
terms being used, which is why the dynamical values of $\chi^{2}_{\nu}$ are all
slightly higher than their kinematic equivalents.)

To date, this situation -- in which Keplerian orbits with a near two-to-one
period ratio provide an excellent fit to the radial velocity data, and yet
correspond to initial conditions that are either dynamically inconsistent
and/or dynamically unstable -- has arisen for three other published RV data sets.
The most dramatic example is the GJ\,876 system \citep{marcy2001} in which
the outer 2.5\,\Mjup\ planet has orbited more than 50 times since the
beginning of observations of the system with the Keck 
telescope \citep{rivera2005}. As has been shown by a number of authors, 
starting with \citet{lc2001}  and \citet{rl2001}, dynamical fits to the GJ\,876
system indicate that the outer two planets are participating in a 2:1 resonance 
in which the critical angles,
$\theta_{1}=2\lambda_{2} - \lambda_{1} - \varpi_{1}$,
and $\theta_{2}=2\lambda_{2} - \lambda_{1} - \varpi_{2}$ 
both librate with small amplitudes of
${\theta_1}_{max}\sim5^{\circ}$ and ${\theta_2}_{max}\sim20^{\circ}$.

The other two stars that appear to harbor planetary companions participating
in the 2:1 resonance are HD\,128311 with $M_1 \sin(i) = 2.6 \, M_{\rm Jup}$,
$M_2 \sin(i) = 3.2 \, M_{\rm Jup}$, $P_1=449 \, {\rm d}$, $P_2=920 \, {\rm d}$,
$K_1=85 \, {\rm m s^{-1}}$, and $K_2=80 \, {\rm m s^{-1}}$ (Vogt et al. 2005), and
HD\,82943, with $M_1 \sin(i) = 1.85 \, M_{\rm Jup}$,
$M_2 \sin(i) = 1.84 \, M_{\rm Jup}$, $P_1=221.6 \, {\rm d}$, $P_2=444 \, {\rm d}$,
$K_1=67 \, {\rm m s^{-1}}$, and $K_2=46 \, {\rm m s^{-1}}$ 
\citep{mayor2004,fm2005,lee2005}. 
While the HD\,73526 system is broadly similar to the HD\,82943
system,  it bears a strikingly close 
outward resemblance to the HD\,128311 system.
Both systems have pairs of near-equal $m \sin(i) \sim 2 M_{\rm Jup}$ companions
and periods in the $P_1 \sim 200\, {\rm d}$, $P_2 \sim 400\, {\rm d}$ range,
though the total radial velocity semi-amplitude is larger
for HD\,73526, and the period is shorter. This means that aside from GJ\,876, 
the HD\,73526 system 
has the best potential for exhibiting non-Keplerian dynamics over time-scales
accessible to radial velocity observations.

\begin{figure}
\centering\includegraphics[angle=-90,width=80mm]{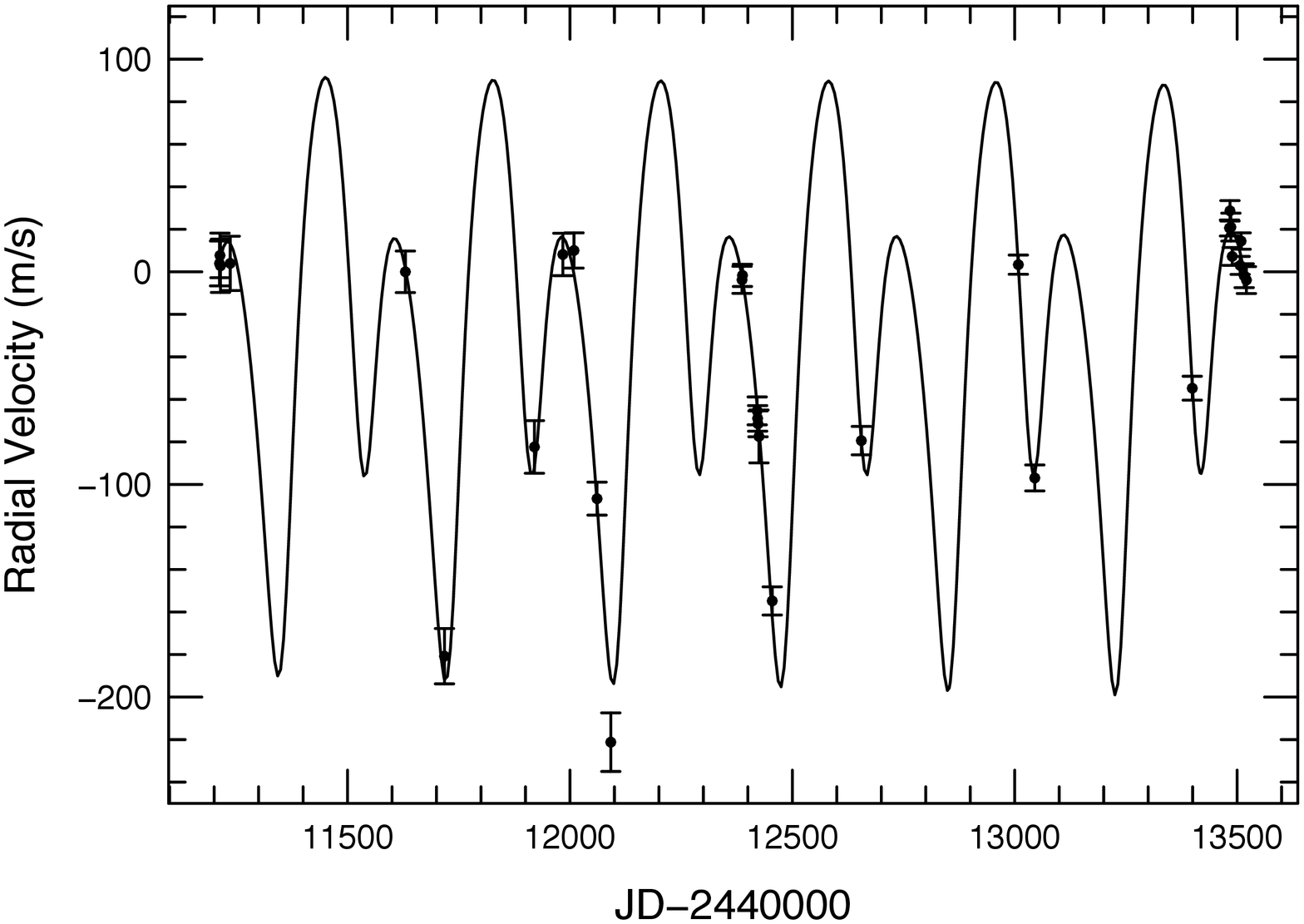}
\caption{Dynamical reflex velocity fit of Table \ref{dyn_orbits}
with measured velocities. The rms residuals to this fit are 7.9\,\ms, 
and the reduced $\chi^2_{\nu}$=1.57. (This fit uses all data epochs and measures
$\chi^2_{\nu}$ relative to the internal uncertainties of Table \ref{vel73526},
which is why the RMS and $\chi^2_{\nu}$ appear higher than those reported
for the kinematic fits.)}
\label{hd73526_dynamic}
\end{figure}

In anticipation of the potential future importance of the HD\,73526 system,
we have carried out a self-consistent 3-body dynamical fit to the 
observed velocities (see \citealt{laughlin2005}). The fitting 
procedure employs the dual Keplerian model listed in Table \ref{kin_orbits} as an initial guess,
and uses a Levenberg-Marquardt algorithm (similar to that described by
\citealt{press}) to obtain osculating orbital elements (defined at 
JD 2451212.1302) that give a stellar reflex velocity that minimizes 
$\chi^{2}$. The orbits are assumed to be co-planar and edge-on. The 
resulting fit is listed in Table \ref{dyn_orbits}, and plotted 
in Figure \ref{hd73526_dynamic}.
\clearpage
\begin{deluxetable}{lcc}
\tablenum{5}
\tablecaption{HD\,73526b,c Dynamical Orbit Fit}
\label{dyn_orbits}
\tablewidth{0pt}
\tablehead{
\colhead{Parameter}   & \colhead{HD\,73526b}
                                                   & \colhead{HD\,73526c}
                                                                           }
\startdata
Orbital period $P$ (d)              & $188.3 \pm 0.9$ & $377.8 \pm 2.4$ \\
Mean Anomaly$^{a}$ ($^{\circ}$)     & $86 \pm 13$     & $82 \pm 27$ \\
Mass (\Mjup)                        & $2.9\pm0.2$     & $2.5\pm0.3$ \\
Eccentricity $e$                    & $0.19 \pm 0.05$ & $0.14 \pm 0.09$\\
$\varpi$                            & $203 \pm 9$     & $13 \pm 76$\\
Velocity offset (\ms)               & \multicolumn{2}{c}{-29.96} \\
Epoch (JD)                          & \multicolumn{2}{c}{2451212.1302} \\
$\chi^{2}_{\nu}$                    & \multicolumn{2}{c}{1.57} \\
RMS (\ms)                           & \multicolumn{2}{c}{7.9} \\
\enddata
\end{deluxetable}
\clearpage
\clearpage
The system implied by this fit is dynamically stable over
a 1\,Myr test integration. It is evidently protected by a 2:1 mean motion
resonance in which $\theta_1$ librates with width 
${\theta_1}_{max}=95^{\circ}$. The resonant argument ${\theta_2}$, however,
is circulating, indicating that the apsidal lines for the orbits precess relative
to one another. The system is subject to the competing influences of both
the 2:1 resonant interaction, which forces $\theta_1$ to librate,
as well as the Laplace-Lagrange secular
interaction, which drives $\varpi_2-\varpi_1$ through the full $2\pi$ range
(see \citealt{md1999}).

In order to derive the quoted uncertainties in the orbital elements of the
dynamical fit, we have
adopted the procedure described in \citet{vogt2005}. We take the
self-consistent 2-planet fit listed in Table \ref{dyn_orbits} and apply a Monte-Carlo
algorithm in which alternate radial velocity data sets are generated
by scrambling the residuals to the fit and then adding them back
with replacement to the model velocities. We then use the 
Levenberg-Marquardt algorithm to generate a self-consistent fit
to each of the Monte Carlo-generated data sets, adopting an unbiased 
initial guess with $P_1=188. \,{\rm d}$, $P_2=377. \, {\rm d}$,
$M_1=M_2=0^{\circ}$, $e_1=e_2=0$, $m_1=2\, M_{\rm Jup}$,
and $m_2=3\, M_{\rm Jup}$. 

When a trial has converged, the resulting system is integrated for 
$10^{4}$\,yr, and the maximum eccentricity attained by each planet during the 
integration is noted. Orbital instability is generally indicated when
the eccentricity of either planet approaches unity. We also monitor the 
maximum excursions of $\theta_1$ and $\theta_2$, in order to determine
whether each individual fit is in 2:1 resonance for the entire $10^{4}$ yr.
In 500 such trials, we obtain 268 unstable systems. Among the remaining
systems, 159 have $\theta_1$ librating and $\theta_2$ circulating, and
49 have {\it both} $\theta_1$ and $\theta_2$ librating. The remaining 24 systems
have both resonant arguments circulating. Integrations of these 24
systems to times
longer than $10^4$ years have shown instability in every case that has been
tested so far.

We note that, as for dual Keplerian fitting, the aliasing is also problematic
for our dynamical fits. This results in a degeneracy between in our ability to determine
$e$ and planet mass.
The resulting uncertainty in these parameters is of similar scale to that seen in
our purely kinematic Keplerian fits. Nonetheless, the period determinations (again as
for our kinematic fits) are well determined, as is the conclusion of 2:1 resonance.

The configuration listed in Table \ref{dyn_orbits}
is strongly reminiscent of the
dynamical fit to the HD\,128311 radial velocity data set 
reported by \citet{vogt2005},
although the orbital eccentricities ($e_{inner}=0.38$, $e_{outer}=0.21$)
were higher in that case than they are here. It will be very
interesting if the planets in
these systems are indeed trapped in dynamical states in which the
2:1 resonance argument $\theta_1$ librates
while $\theta_2$ circulates. Librating-circulating
configurations are not observed to arise from the coplanar migration
scenarios that have been studied by Lee and collaborators 
\citep{lp2002,lee2004,lee2005}, and which have been successfully 
applied to model the origin of the GJ\,876 and HD\,82943 systems.
In the Lee et al. migration scenarios, the planets invariably wind up having
both arguments in libration. Configurations such as the one given
in Table \ref{dyn_orbits} would have to arise either through (1) migration with initial 
planetary eccentricities,
(2) via migration that occurred very rapidly, or (3) as a result of 
a dynamical scattering event. Work is currently underway to study these
three possibilities.

The unusual best-fit dynamical configuration for HD 73526, in which the 
planets participate in 2:1 resonance with only $\theta_1$ librating, 
is apparently not an accessible state of the co-planar disk migration 
scenarios investigated by \citet{lee2004}. It may be possible, however, 
for capture into a $\theta_1$-librating, $\theta_2$-circulating state to 
occur via fast migration, or migration with initial eccentricities, or migration 
with with significant initial mutual inclination (Lee, M.H., 
private communication). In theory, the 
dynamical interactions between planets ``b'' and ``c'' should allow the 
mutual inclination of the planets to be obtained via dynamical fitting 
to radial velocities. In practice, however, such a determination will be 
difficult to carry out. Even the much more extensive, higher signal-to-noise 
GJ 876 data set is not yet sufficient to allow a definitive measurement of 
mutual inclination \citep{rivera2005}.

It has been suggested by \citet{gk2005} that the
radial velocity variations in the HD\,128311 and HD\,82943 systems 
are caused not by 2:1 resonant configurations, but rather by pairs of planets
in 1:1 resonances that resemble high-eccentricity retrograde
satellite configurations (see \citealt{lc2001}).
We have searched for dynamical fits to the HD\,73526 data set 
which involve pairs of planets
in 1:1 resonance, but were not able to find satisfactory co-planar,
$i=90^{\circ}$ fits. The dearth of such fits 
likely arises from the fact that the 
osculating orbital eccentricities of the HD\,73526 planets are smaller
than in either the HD\,128311 or HD\,82943 systems.

\section{Conclusion}

The HD\,73526 resonant system joins a growing list of exoplanetary multiples in
resonant configurations -- if it is added to the seventeen multiple planets
published or in press as at September 2005 \citep{marcy05a}, we have 8 of 18
planetary systems containing at  least one resonance, and 4 of 18 planetary systems
(Gl\,876, HD\,82943, HD\,1128311 and HD\,73526) containing a 2:1 resonance. Given
the difficulties imposed by the detection of multiple planetary systems (it takes
many more observations to detect a multiple than it does to detect the largest
velocity signature in a system), and the tendency for resonances to be masked by
aliasing and window function effects, these numbers are almost certainly lower
limits.

The core-accretion paradigm for planetary formation predicts gas
giant planets to form in circular orbits beyond 3-5\,AU.  However, planets
orbiting beyond 0.1\,AU (i.e. not circularised  by the host star) are seen to have a
median eccentricity of 0.25 \citep{marcy05a}, so exoplanets in circular orbits are
the exception, rather than the rule. Interactions  between forming planets and
their gaseous disk are thought to dampen,  rather than excite, eccentricities
\citep{tw2004}. This suggests that orbital eccentricities arise {\em after} 
major gas accretion, and that the observed orbital
eccentricities are the result of subsequent  interactions  between  planets and their
disk, or between planets as they migrate. Certainly, the significant number of 
gas giants found on small
orbits, suggests that orbital migration for such gas giants must be common, 
if not almost ubiquitous.
If this is indeed the case, then the detection of a large number of planetary
systems trapped into stabilising resonances (more than a third at present) 
would arise as a logical consequence.


\acknowledgments
The Anglo-Australian Planet Search team would like to gratefully 
acknowledge the superb technical support which has been
received throughout the programme from AAT staff - in particular R.Patterson, 
D. Stafford, J. Collins, S. Lee, J. Pogson, G. Kitley, J. Stevenson, and S. James. 
We further acknowledge
support by; the partners of the Anglo-Australian Telescope Agreement
(CGT,JAB,BDC,HRAJ); NASA grant NAG5-8299 \& NSF grant AST95-20443 (GWM);
NSF grant AST-9988087 (RPB); NASA grant NNG04G191G (GL).  NSO/Kitt Peak
FTS data used here were produced by NSF/NOAO.  This research has
made use of the SIMBAD database, operated at CDS, Strasbourg, France,
and the NASA's Astrophysics Data System.


\end{document}